\documentclass[
reprint,
superscriptaddress,
floatfix,
 amsmath,amssymb,amsfonts,
aps,
prb,
]{revtex4-2}
\usepackage{graphicx}
\usepackage{dcolumn}% Align table columns on decimal point
\usepackage{verbatim}
\usepackage{bm}% bold math
\usepackage{color}
\usepackage{hyperref}% 
\usepackage[dvipsnames]{xcolor}
\hypersetup{
    colorlinks=true,
    linkcolor=Blue,
    citecolor=Blue,   
   urlcolor=Blue,
   }

\def\bea{\begin{eqnarray}}
\def\eea{\end{eqnarray}}
\def\beq{\begin{equation}}
\def\eeq{\end{equation}}
\def\f{\frac}

\def\be{\beta}
\def\D{\Delta}

\def\a{\alpha}

\def\r{\rho}

\def\la{\langle}
\def\ra{\rangle}

\def\d{\delta}

\def\a{\alpha}
\def\d{\delta}
 
\def\la{\langle}
\def\ra{\rangle}

\def\bea{\begin{eqnarray}}
\def\eea{\end{eqnarray}}

\def\r{\rho}
\def\dr{\delta \rho}
\def\dc{\delta c}

\def\d{\partial}
\def\gr{\mathbf{\nabla}}

\def\dt{\cdot}

\def\rb{\mathbf{r}}

\def\b{\beta}
\def\r{\rho}
\def\dr{\delta \rho}
\def\dc{\delta c}

\def\d{\partial}
\def\gr{\mathbf{\nabla}}

\def\dt{\cdot}

\def\rb{\mathbf{r}}

\def\A{\mathcal{A}}
\def\B{\mathcal{B}}
\def\C{\mathcal{C}}
\def\D{\mathcal{D}}
\begin{document}
\title{Self-organized fractal architectures driven by motility-dependent chemotactic feedback}
\author{Subhashree Subhrasmita Khuntia}
\affiliation{Department of Physical Sciences, Indian Institute of Science Education and Research Mohali, Sector 81, Knowledge City, S. A. S. Nagar, Manauli PO 140306, India}

\author{Debasish Chaudhuri}
\email[For correspondence:~]{debc@iopb.res.in}
\affiliation{Institute of Physics, Sachivalaya Marg, Bhubaneswar 751005, India}
\affiliation{Homi Bhabha National Institute, Anushakti Nagar, Mumbai 400094, India}

\author{Abhishek Chaudhuri}
\email[For correspondence:~]{abhishek@iisermohali.ac.in}
\affiliation{Department of Physical Sciences, Indian Institute of Science Education and Research Mohali, Sector 81, Knowledge City, S. A. S. Nagar, Manauli PO 140306, India}

\date{\today}

\begin{abstract}
Complex spatial patterns in biological systems often arise through self-organization without a central coordination, guided by local interactions and chemical signaling. In this study, we explore how motility-dependent chemical deposition and concentration-sensitive feedback can give rise to fractal-like networks, using a minimal agent-based model. Agents deposit chemicals only while moving, and their future motion is biased by local chemical gradients. This interaction generates a rich variety of self-organized structures resembling those seen in processes like early vasculogenesis and epithelial cell dispersal.
We identify a diverse phase diagram governed by the rates of chemical deposition and decay, revealing transitions from uniform distributions to sparse and dense networks, and ultimately to full phase separation. At low chemical decay rates, agents form stable, system-spanning networks; further reduction leads to re-entry into a uniform state. A continuum model capturing the co-evolution of agent density and chemical fields confirms these transitions and reveals how linear stability criteria determine the observed phases.
At low chemical concentrations, diffusion dominates and promotes fractal growth, while higher concentrations favor nucleation and compact clustering. These findings unify a range of biological phenomena—such as chemotaxis, tissue remodeling, and self-generated gradient navigation—within a simple, physically grounded framework. Our results also offer insights into designing artificial systems with emergent collective behavior, including robotic swarms or synthetic active matter.
\end{abstract}

%\keywords{Auto-chemotaxis $|$ Concentration-dependent sensitivity $|$ Motility controlled deposition $|$ fractal-network architecture}

\maketitle

\section{Introduction}

The spatial architecture of living systems emerges through self-organization without a central coordination~\cite{Bonabeau2001, Camazine1997, Prisca2018}, often, with information propagating via self-generated chemical signals %evolving within the physical milieu to a steady state 
\cite{Zhang2019, Levin2013, Lewis2019, Whitchurch2015}. 
These systems display large-scale patterns, from shell and neural structures \cite{Schmahl2021, Oster2009, Faber2022, Zapotocky2011} to ecological distributions \cite{Korman2017, Li2006}. Communication among components is crucial for such coordinated patterning, enabling adaptive responses to environmental shifts \cite{Bassler2012, Bassler2019}. Signal dynamics and adaptation, in turn, limit communication efficacy \cite{Ping2020}. A key aspect of self-organization is the modification of the environment by motile agents through localized signals, influencing subsequent behaviors \cite{Golestanian2019, Vicsek2014}. This manifests in quorum sensing, biofilm formation, ant trails, neural fasciculation, and pedestrian tracks, generating spatiotemporal structures that reflect adaptive functionality and reveal underlying local rules \cite{Lowen2018, Sourjik2016, Giuggioli2018, Theraulaz2016, Dussutour2004, Breau2023, Chaudhuri2011, Helbing1997}.

Such self-organisation should, in principle, extend to eukaryotic chemotaxis, essential for wound healing, development, and cancer metastasis~\cite{stuelten2018cell,insall2013interaction}. However, predicting cellular chemotactic responses remains challenging due to insufficient information about chemoattractant sources and their interaction with the cell environments~\cite{insall2022steering}. Often seen as a passive process where cells follow external gradients, chemotaxis fails to account for the active role cells play in shaping these patterns~\cite{dona2013directional,tweedy2020seeing,tweedy2016self, Alert2022}, thus overlooking important physiological mechanisms. For instance, cells like melanoma and Dictyostelium not only respond to chemoattractants but also actively degrade them (e.g., lysophosphatidic acid and cAMP), modulating the very gradients they navigate~\cite{muinonen2014melanoma,sucgang1997null}. Similarly, epithelial cells at low densities form branched, network-like structures in response to growth factor limitation~\cite{leggett2019motility}, suggestive of an active, self-patterning mechanism. Such structures are further seen during the early stages of vasculogenesis in embryonic development~\cite{risau1995vasculogenesis,szabo2008multicellular,gamba2003percolation,merks2006cell,merks2008contact}. Despite these observations, the minimal ingredients necessary to produce such structures—particularly fractal-like networks—remain unclear. In this study, we explore how such behavior-mediated changes can give rise to fractal network-like structures within a chemotactic framework. 
Despite the distinct physiological properties and underlying causes of these processes, our findings show that similar structures can emerge naturally from chemotactic behaviours, providing a unified framework for understanding the development of spatial patterns in living systems.

We investigate an agent-based model where agents act as  excluded volume diffusers depositing chemicals along their paths. 
%persistent random walkers with negligible persistence on the scale of their body length,
These chemical trails influence their subsequent movements, guiding them towards areas of higher chemical concentration~\cite{Drury2020, VanBrussel2004}. In natural systems, however, the sensitivity of agents to concentration changes is typically limited. To account for this, we incorporate concentration-dependent sensitivity into our model~\cite{Romanczuk2014}. As the chemical concentration exceeds a critical threshold, agents’ bias toward higher concentrations plateaus. Chemical dynamics in the model include both evaporation and deposition processes. Evaporation is stochastic with a decay rate $\beta$, while deposition, characterized by $\a$ and local chemical activity, depends on motility to optimize resource utilization~\cite{Ratnieks2013, Merks2020}. Specifically, agents release chemicals solely in motion, ensuring resource efficiency, ceasing when movement halts due to crowding-induced repulsion. These factors, concentration-limited responsiveness and movement-dependent deposition, strongly influence the system's self-organization.

\begin{figure*}[htbp]
    \centering
    \includegraphics[width=0.9\linewidth]{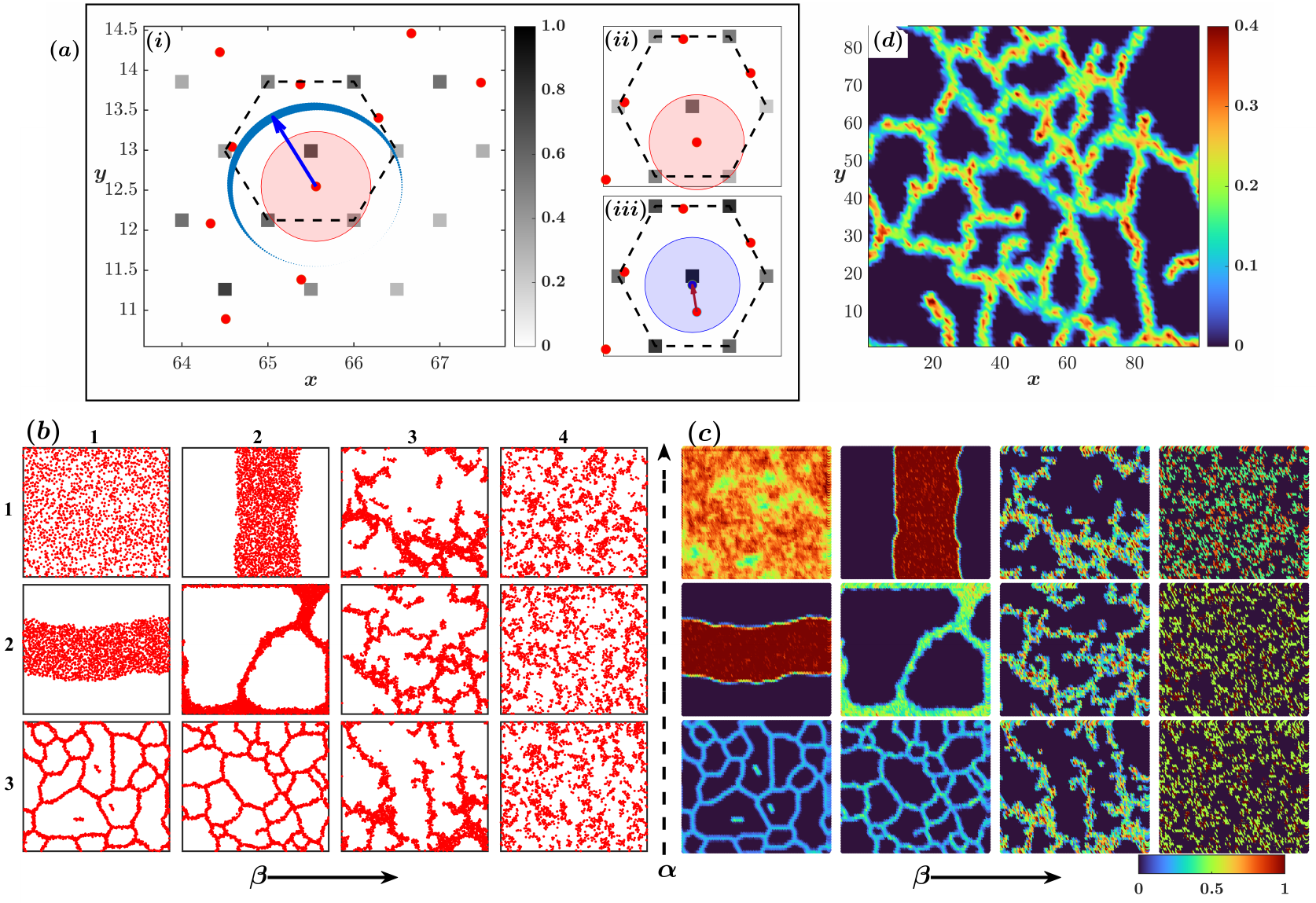}
    \caption{
 (a)~Schematic of agent-based rules: (a-i)~A focal particle (red-shaded circle) senses chemical concentration from surrounding grid points (gray boxes) within a hexagonal neighborhood. The movement direction (blue arrow) is determined by a weighted sum of local chemical gradients and Gaussian noise (width $\pi/4$). (a-ii)~If a trial move is rejected, no chemical is deposited. (a-iii)~If accepted, the particle moves and deposits chemical at the new location (darker gray grid). Move lengths are drawn uniformly from $[0, \sigma]$.
(b, c)~Particle configurations (b) and chemical profiles (c) shown across $\alpha$–$\beta$ space. 
Rows 1–3 correspond to $\alpha = 1, 50, 100$, while columns 2–4 have $\beta = 0.008, 0.1, 0.5$. Column 1 uses $\beta = 0.002, 0.004, 0.0005$ for rows 1–3, respectively. At high $\beta$, transient micro-clusters form in the homogeneous phase (HP, column 4, figures $b_{14}$, $b_{24}$, $b_{34}$).  Lower $\beta$ yields sparse network (SN, column 3, figures $b_{13}$, $b_{23}$, $b_{33}$) dense network (DN, $b_{31}$, $b_{32}$, $b_{22}$), single clusters (SC, $b_{21}$, $b_{22}$), and re-entrant homogeneous phase (RHP, $b_{11}$).  The colorbar represents the concentration of the chemical in units of $\frac{\a}{\be}$ (d)~Chemical profile at $\alpha = 5.0$, $\beta = 0.008$ shows efficient, system-spanning fractal clusters with low chemical cost.
    }
    \label{fig:scheme}
\end{figure*}

\section{Results}

The details of the model is outlined in the Methods section with the help of Fig.\ref{fig:scheme}(a). 
Consider a uniform distribution of $N$ excluded-volume agents, moving randomly like particles in a gas. In the presence of a significant self-generated chemical landscape, agents are drawn to areas of higher chemical concentration through a positive feedback loop.
Over time, this feedback loop gives rise to spatial inhomogeneities in both agent distribution and chemical concentration. As the chemical field strengthens, agents are increasingly drawn toward regions of higher concentration, leading to clustering instabilities reminiscent of the Keller-Segel (KS) model of chemotaxis. In the KS framework, such instabilities typically emerge when chemical sensitivity or local concentration exceeds a critical threshold.

To explore how this feedback mechanism governs emergent structure, we systematically vary the chemical deposition rate ($\alpha$) and decay rate ($\beta$) and examine the resulting morphologies.
\begin{figure*}[t!]
    \centering
    \includegraphics[width=\linewidth]{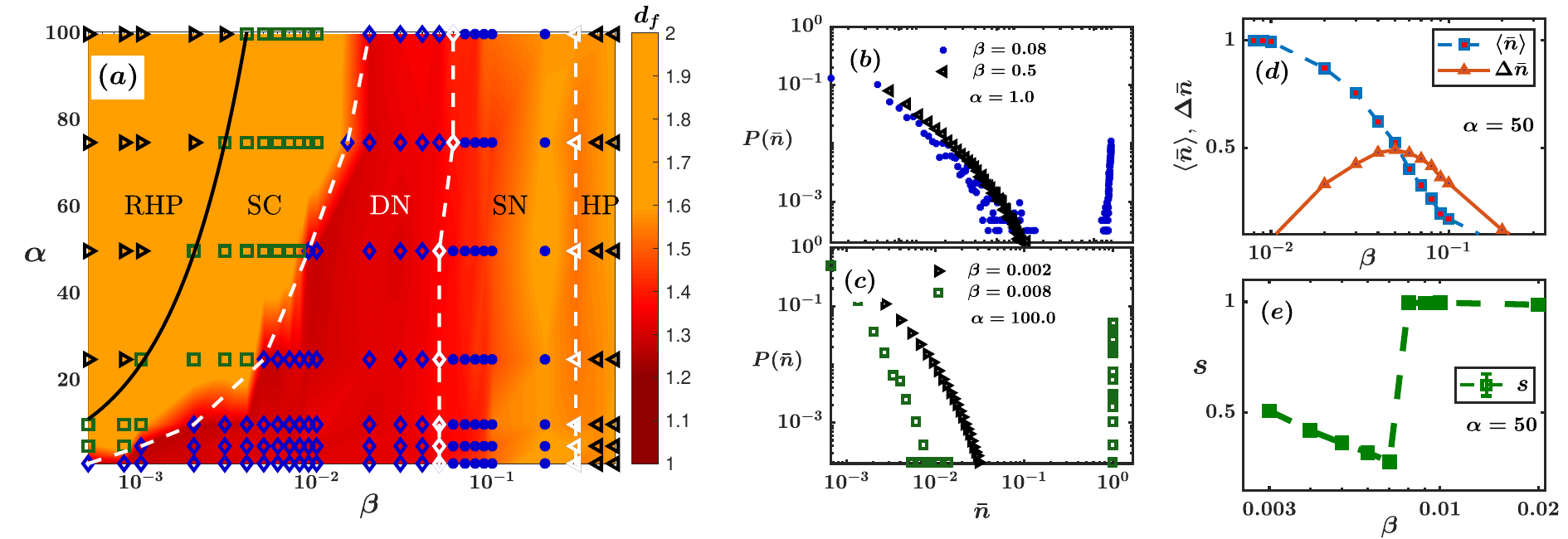}
    \caption{  
   (a)~Phase diagram showing structural transitions with $\alpha$ and $\beta$. Background indicates fractal dimension $d_f$. At high $\beta$, particles form a homogeneous phase (HP, $\triangleleft$) with no stable clusters (b). A second homogeneous phase (RHP, $\triangleright$) appears at lowest $\beta$ (c). Aggregates emerge in between, separated from RHP by a linear stability boundary obtained using ~Eq. \ref{eq::instability} (solid line).  Decreasing $\beta$ from HP leads to a sparse network (SN, $\bullet$), then a fractal network (DN, $\diamond$), and finally dense clusters (SC, $\square$); dashed lines mark phase boundaries. (b, c)~Cluster size distributions $P(\bar{n})$ at selected $\alpha$, $\beta$ values. (d)~Mean cluster size and its standard deviation, with a peak in the deviation marking the SN–DN transition. (e)~A drop in the area spanned by clusters $s$ signals the DN–SC transition; $s$ rises at low $\beta$ as SC approaches the RHP phase.
   }
    \label{fig:Phasediagram} 
\end{figure*}

\subsection{Re-entrant transition and system spanning network}
The morphology of the system evolves dramatically with changing chemical parameters, illustrated through typical particle configurations and the chemical profile in Fig. \ref{fig:scheme}(b) and (c) respectively. At high decay rates (large $\beta$), rapid evaporation prevents long-range chemical accumulation, and agents remain homogeneously distributed (HP phase) with minor, transient clustering (fourth columns of Fig. \ref{fig:scheme}(b) and (c), Supplementary Movie S1). As $\beta$ decreases, the slower decay allows local concentrations to build up, reinforcing aggregation via chemotactic feedback. This initiates the formation of sparse networks (SN phase, third column of Fig. \ref{fig:scheme}(b) and (c), Supplementary Movie S2), which transition into dense, branching structures as decay is further reduced (columns 1 and 2 in row 3 of Fig. \ref{fig:scheme}(b) and (c), Supplementary Movie S3). 
At low $\be$ and moderate $\a$, sharp chemical gradients develop, stabilizing intricate, system-spanning fractal networks (Fig. \ref{fig:scheme} d). These networks are stabilized by directional chemical bias and spatial exclusion, which prevent collapse into compact clusters. At higher deposition rates ($\alpha$), the network structures collapse into a large, dense aggregate marked as a single cluster phase (SC, row 1, column 2 and row 2, column 1 in Fig. \ref{fig:scheme}(b,c)  Supplementary Movie S4). However, at very low $\be$, the system transitions again: chemical buildup becomes so strong that agent sensitivity saturates, weakening the chemotactic bias, resulting in a re-entrant homogeneous phase (RHP, Supplementary Movie S1), where particles become effectively insensitive to the uniform chemical field (Fig. \ref{fig:scheme}(b),(c), First row, first column).
Thus, the model reveals a non-monotonic transition as $\be$ decreases: the system moves from a disordered homogeneous phase to networks and clusters, and then returns to a second homogeneous state at extreme parameter values. This re-entrant behavior arises not from the depletion of signals, but from the loss of responsiveness due to oversaturation — an effect particularly relevant in biological systems where chemotactic sensitivity is dynamically regulated.

\subsection{Phase diagram}

Figure~\ref{fig:Phasediagram}(a) presents the phase diagram and a heatmap of the fractal dimension \( d_f \), computed via the correlation dimension method \cite{Grassberger1983,strogatz} (see SI, Fig.~S5). Lower \( d_f \) indicates network-like states, while \( d_f \approx 2 \) corresponds to homogeneous or compact cluster phases. Phase boundaries, shown as visual guides, separate five regimes (right to left): homogeneous phase (HP), sparse network (SN), dense network (DN), system-spanning cluster (SC), and re-entrant homogeneous phase (RHP). Representative points are evaluated and described in the diagram. The visually distinct phases in particle configurations and chemical profiles [Fig.~\ref{fig:scheme}(b,c)] are quantitatively characterized in Sec.~\ref{sec:cluster}. The instability in homogeneous phase  is analyzed via mean-field dynamics in Sec.~\ref{sec:LSA}.

\subsection{Cluster size}
\label{sec:cluster}

We analyze the scaled cluster size distribution $P(\bar{n})$, where $\bar{n} = n/N$, to distinguish homogeneous (HP \& RHP) from clustered states (SN \& SC), using a clustering algorithm with cutoff separation $r_c = 2\sigma$.  
Figure~\ref{fig:Phasediagram}(b) distinguishes the SN phase from the HP phase. In SN, $P(\bar{n})$ is bimodal — an approximately exponential decay for small clusters coexists with a delta-like peak near $\bar{n}=1$, indicating an infinite cluster. In contrast, the HP phase shows an approximately homogeneous particle distribution, with $P(\bar{n}) \sim \bar{n}^{-\nu} \exp(-\bar{n}/\bar{n}^*)$~\cite{Majumdar1998,Peruani2013}, where $\nu \approx 3/2$ and $\bar{n}^*$ is the typical cluster size (see SI, Fig.~S3, and Supplementary Movie S1). The power-law correction reflects giant number fluctuations (GNF) due to transient microclusters (see SI, Fig.~S4). A similar $P(\bar{n})$ signature marks the re-entrant SC–RHP transition in Fig.~\ref{fig:Phasediagram}(c).

To identify transitions between clustered phases (SN and DN), we use the mean cluster size $\langle \bar{n} \rangle$ and its standard deviation, $\Delta\bar{n}$ (Fig.~\ref{fig:Phasediagram}(d)). As $\beta$ decreases within an intermediate range, $\langle \bar{n} \rangle$ rises and saturates at 1, indicating the emergence of a system-spanning cluster. Notably, $\Delta\bar{n}$ peaks near the SN–DN transition. In addition to fractal dimension, $d_f$, DN and SC phases are further distinguished by the scaled convex hull area $s$ (cluster area relative to the simulation box; see SI, Fig.~S6). In Fig.~\ref{fig:Phasediagram}(e), $s$ drops sharply from 1 as $\beta$ decreases, signaling the DN–SC transition. With further decrease in $\beta$, $s$ rises again as the spanning cluster grows.

\begin{figure*}[ht]
    \centering
    \includegraphics[width=\textwidth]{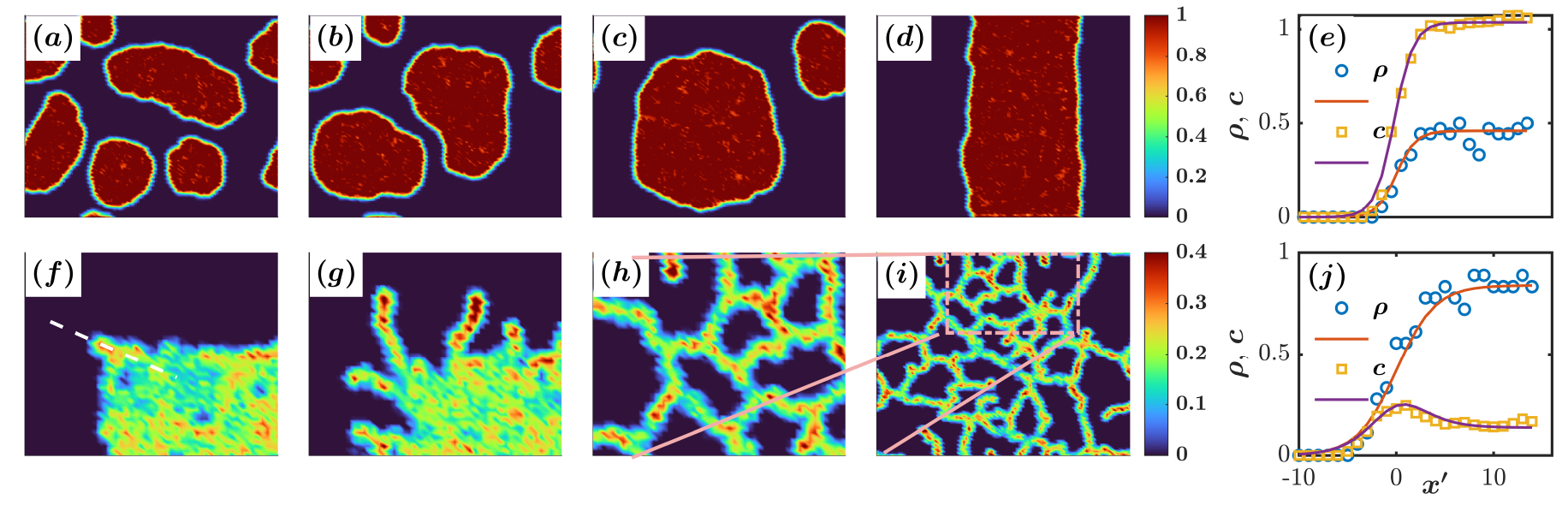}
    \caption{Typical configurations during quench from a homogeneous to SC phase at $\alpha = 100.0$, $\beta = 0.008$ are shown in (a–d) at times $t = 8 \times 10^4$, $1.6 \times 10^5$, $8 \times 10^5$, and $3 \times 10^6$. The lower panel (f–i) shows the evolution from a compact initial state at $\alpha = 5.0$, $\beta = 0.008$ corresponding to DN phase, over times $t = 10^3$, $10^4$, $10^5$, and $10^5$.
    (j) shows particle and chemical density profiles, $\rho$ and $c$, along the dashed diagonal in (f). (e) presents similar profiles for a configuration like (f), but at $\alpha = 100.0$, $\beta = 0.008$, corresponding to the SC phase in the upper panel (Supplementary movie S5). The fitted lines follow $\r(x)=\frac{\r_i}{2}[1+\tanh(x/w)]$  and $c(x) = \r(x)[\r_c-\r(x)]$ in units of $\a/\be$, where $\r_i=0.46$, $w=2.0$, $\r_c=2.712$ in (e) and $\r_i=0.84$, $w=4.0$, $\r_c=1.002$ in (j), respectively. 
    }
    \label{fig:perturb}
\end{figure*}

\subsection{Continuum Description and Linear Stability Analysis}
\label{sec:LSA}
In this section, we develop an effective continuum model starting from the microscopic dynamics. The time evolution of the $k$-th particle at position $\rb_k$ coupled to the  chemical field $c(r,t)$, is described by:
%\alert
{
\bea 
\nonumber
\mathbf{\dot{r}}_k &=&  -\mu \gr_k U + \mu \gr_k f(c) + \sqrt{2D_T}~ \mathbf{\eta}_k (t) \\
\dot{c} &=& \a h(\{\rb, \dot{\rb}\}) - \beta c
\label{eq::position}
\eea
}

Here, $U = \sum_{k<k'} u(|\mathbf{r}_k- \mathbf{r}_k'|) $ represents the exclusion interaction between the walkers, modeled by the Weeks-Chandler-Andersen (WCA) potential~\cite{Weeks1971, Chandler1983} in the simulation. The function $f(c) = c/(1+c)$ defines how the particle motion is influenced by the local chemical concentration. Translational diffusion, with coefficient $D_T$, is driven by uncorrelated Gaussian noise satisfying $\langle \eta_k(t) \rangle = 0$ and $\langle \eta_k(t)\eta_{k'}(t') \rangle = \delta_{kk'}\delta(t - t')$. The motility-dependent deposition rate $h(\{\rb, \dot{\rb}\})$ decreases in dense regions where volume exclusion hinders particle motion. 

%\alert
{In the absence of chemical coupling, local repulsion in an excluded volume system can be approximated as $V_{\rm int} = \int d {\bf x} \, d {\bf x'} \, \r({\bf x}) \lambda \delta({\bf x} - {\bf x'}) \r({\bf x'})$, enhancing collective diffusivity to $D = D_T + 2\lambda$. The coupled evolution of particle density and chemical concentration is:
\bea
\nonumber
    \dot{\rho} &=& -\gr \dt[\r \chi f'(c) \gr c-D \nabla \rho] \\
    \dot{c} &=& \alpha \rho (\r_c-\r)-\beta c . 
    \label{eq::density}
\eea
The functional dependence $\r(\r-\r_c)$ captures the decrease in the rate of chemical deposition with reduced motility at high density controlled by $\r_c$ (see SI; Fig.S1 and S2).

We conduct a linear stability analysis around the homogeneous state $(\r_0, c_0)$, where $c_0 = \r_0(\r_c - \r_0) \frac{\a}{\be}$. Since the trace of the stability matrix satisfies $Tr({\cal M}) < 0$, instability arises when $Det({\cal M}) = 0$, resulting in the quadratic equation $c_0^2 + (2 - \chi')c_0 + 1 = 0$, with $\chi' = \frac{(\r_c - 2 \r_0) \chi}{(\r_c - \r_0) D}$. The solutions $ C_\pm = \frac{\chi' - 2 \pm \sqrt{\chi'^2 - 4\chi'}}{2}$ define the range of $\a/\be$ values within which the homogeneous state becomes unstable towards pattern formation:
\bea
 \f{ C_-}{\r_0 (\r_c-\r_0)} < \frac{\a}{\be} < \f{C_+}{\r_0 (\r_c-\r_0)}. 
 \label{eq::instability}
\eea
This condition aligns with the observed pattern formation at intermediate $\beta$ for a fixed $\alpha$, including the re-entrance to the homogeneous phase in Fig.~\ref{fig:Phasediagram}(a) and the KS instability.

At low chemical concentrations, diffusion dominates over attraction, leading to a homogeneous phase. As concentration increases, particles begin clustering. This supports the first bound conceptually. However, unlike the linear stability analysis (Eq. \ref{eq::instability}), the transition boundary in our simulations is independent of $\alpha$.
%This can be explained in the following way:\\

As walkers begin to cluster, only those at the edges deposit chemical. At high $\beta$, rapid evaporation weakens chemical binding, causing the clusters to quickly disperse. This explains the transient clustering and giant number fluctuations observed in HP (see SI). The second bound, shown in Fig.~\ref{fig:Phasediagram}a, marks the phase boundary between the RHP and SC phases at very high chemical concentrations. However, the KS instability does not directly capture the fractal nature of the DN phase.

\subsection{Quench Dynamics}

To explore the dynamics leading to the SC and DN phases, we perform two quench simulations (Supplementary movie S5). Quenching to the SC phase from a homogeneous state shows nucleation and growth (Fig.~\ref{fig:perturb}(a–d)), with mean cluster size scaling as $\langle \bar n \rangle \sim t^{1/3}$ (see SI, Fig.~S7). In contrast, quenching to the DN phase from a compact state yields a network structure. The compact distribution in Fig.~\ref{fig:perturb}(f) shows a density profile along the dashed diagonal as $\rho(x) = \frac{\rho_i
}{2}[1+\tanh(x/w)]$, with $\rho_i$ as the interior density and $w$ the interface width. In steady state, the corresponding chemical profile is $c(x) = \rho(x)[\rho_c - \rho(x)]$ (in units of $\alpha/\beta$), peaking near the interface and obeying  $\rho_c < 2\rho_i$; see Fig.~\ref{fig:perturb}(j). The diffusive flux is maximal along the diagonal, with a value $\sqrt{2}$ times that along the $x$ or $y$ axes. At low chemical deposition—as in Fig.~\ref{fig:perturb}(f–i)—this flux dominates early dynamics, ejecting particles along the diagonal. The resulting chemical peak stabilizes the emerging thin layer, forming the first finger-like pattern. This process repeats at new corners, progressively generating a dense network~(see SI; Fig.~S1 and S8). 

At higher chemical deposition rates ($\rho_c > 2\rho_i$), the local peak in $c(x)$ vanishes (Fig.~\ref{fig:perturb}(e)), and the chemical closely tracks the density, enhancing its stability. This regime aligns with the nucleation-growth behavior in Fig.~\ref{fig:perturb}(a–d), where the DN instability is absent, yielding steady compact bands as in Fig.~\ref{fig:perturb}(d).

\begin{figure*}[t]
    \centering
    \includegraphics[width=\textwidth]{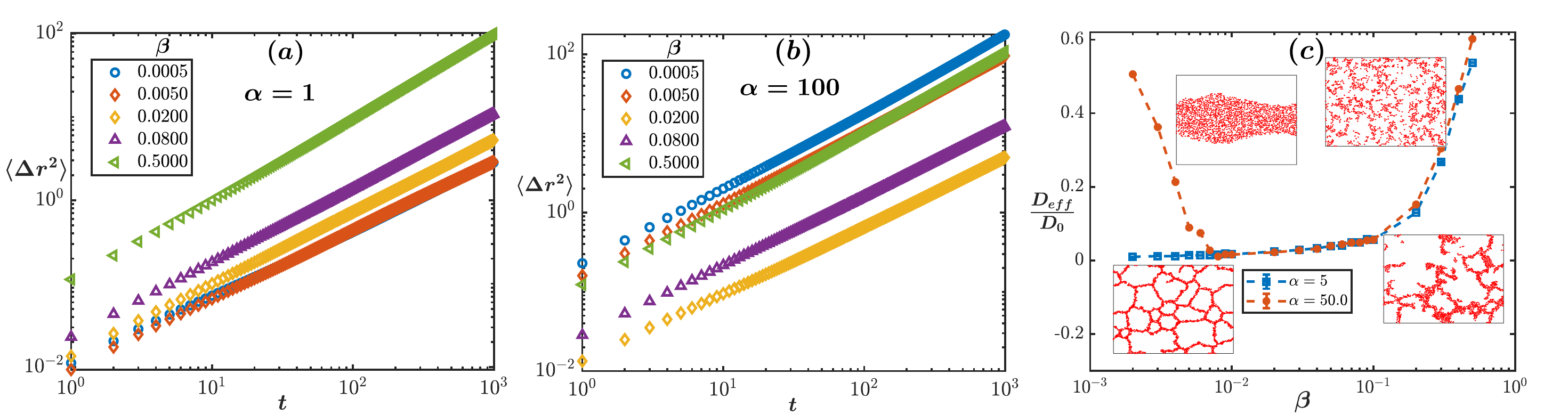}
    \caption{
    (a–b)~Log-log plots of mean squared displacement for $\alpha = 1$ and $100$ at various $\beta$ values. (c)~Effective diffusivity $D_{\text{eff}}$, scaled by the bare diffusivity $D_0$ (at $\alpha = 0$), versus $\beta$ for $\alpha = 5$ and $50$. 
       }
    \label{fig:MSD}
\end{figure*}

\subsection{Mean squared displacement}
The evolution of network structures is linked to walker dynamics. Here, we focus on the mean squared displacement (MSD) of each walker, averaged over all particles (Fig.~\ref{fig:MSD}). The MSD exhibits diffusive scaling at long times across all cases, though the slope in the diffusive regime — indicating the walker’s diffusivity — varies with different parameter sets. Additionally, we observe non-monotonic behaviour in the MSD for high deposition (Fig.~\ref{fig:MSD}(b)). The effective diffusivity, $D_{\text{eff}}$, is shown in Fig.~\ref{fig:MSD}(c) for two $\alpha$ values as $\beta$ changes. For the lower deposition rate ($\alpha = 5$ in Fig.~\ref{fig:MSD}(c)), $D_{\text{eff}}$ decreases as $\beta$ decreases and asymptotically approaches zero. Reduced evaporation increases the chemical bias, leading to walker aggregation and low diffusivity. 
At high deposition ($\alpha = 50$), MSD and $D_{\text{eff}}$ vary non-monotonically with $\beta$: diffusivity first decreases, then rises as $\beta$ is further reduced. This re-entrant behavior is due to the reduced chemical response at higher concentrations, enabling more random movement.

These trends align with the phase diagram (Fig.~\ref{fig:Phasediagram}): at low $\alpha$, $D_{\text{eff}}$ decreases from HP to DN as $\beta$ drops. At high $\alpha$, a re-entrant transition appears — $D_{\text{eff}}$ falls in DN, rises in SC, and reaches high values in RHP at very low $\beta$, resembling the HP phase.

\section{Discussion}

We have presented a minimal agent-based model to explore how motility-dependent chemical deposition and concentration-limited responsiveness can lead to complex self-organized patterns in active systems. By incorporating local interactions, feedback between movement and chemical signaling, and excluded volume effects, our model generates a rich phase diagram, including transitions between homogeneous states, network-like morphologies, dense clusters, and re-entrant homogeneity.

A central outcome of our model is the emergence of fractal, system-spanning networks that closely resemble patterns seen in biological processes such as vasculogenesis and epithelial tissue remodelling. These networks appear when agents deposit chemicals only while moving and respond to chemical gradients with a saturating sensitivity. At intermediate values of chemical deposition and decay, agents form extended, branching structures that are stable yet highly dynamic. At very low decay rates, chemical accumulation drives full phase separation and large aggregate formation. Interestingly, a re-entrant transition back to a homogeneous state is observed at extremely low decay rates, where the sensitivity of agents becomes effectively saturated.

This framework connects well with observations in vasculogenesis and angiogenesis~\cite{carmeliet2000angiogenesis,risau1995vasculogenesis} mediating blood vessel formation, essential for wound healing and tumour growth. In vasculogenesis, endothelial cells (ECs), which line blood vessels, organise into solid cords that remodel into a vascular network. Previous studies have attributed vascular network formation to contact inhibition in cellular Potts models, where cell compressibility leads to effective pressure within aggregates migrating toward a chemoattractant~\cite{merks2008contact,szabo2008multicellular,gamba2003percolation,merks2006cell}. When cultured on Matrigel, a mimic of the extracellular matrix, ECs form honeycomb-like patterns with cords surrounding empty spaces, suggesting that ECs have intrinsic patterning abilities that allow them to organize without relying on external morphogen cues. They produce and respond to various chemoattractants such as VEGF-A, SDF-1, FGF-2, Slit-2, and semaphorins, which regulate their behaviour during vascular development~\cite{helmlinger2000formation,coultas2005endothelial}. ECs also express VE-cadherin, which modulates their motility and cohesion~\cite{gory1999role,perryn2008vascular}.  The reduced motility of ECs in response to high local chemoattractant concentrations may mimic the mechanisms captured in our model, leading to spontaneous organization into vascular-like structures.

Similar principles may apply to epithelial-mesenchymal transition (EMT), where epithelial cells break contact and migrate individually or in groups~\cite{mehes2014collective, wong2014collective}. In vitro, epithelial cell clusters disperse in response to growth factors like EGF (epidermal growth factor) or HGF (hepatocyte growth factor)~\cite{pope2008automated}. A partial EMT generates leader cells with enhanced motility that guides follower cells while maintaining some cell-cell adhesion~\cite{omelchenko2003rho}. At low densities and reduced EGF, non-transformed mammary epithelial cells  form multicellular clusters with branched, fractal-like morphologies~\cite{leggett2019motility}, which only occur when proliferation and motility are suppressed by growth factor limitation. Interestingly, the EGF receptors transmit stimulatory signals to the cytoplasm and also endocytose and degrade ligand EGF, enabling breast cancer cells to migrate by generating self-created EGF gradients~\cite{scherber2012epithelial}. Our results suggest that the observed fractal-like architecture may be explained by motility-limited aggregation, where chemical feedback stabilizes low-density, highly branched morphologies. While previous explanations relied on analogies to non-living systems (e.g., diffusion-limited aggregation), our model offers a biologically plausible mechanism grounded in active motility and local signal degradation.

Chemotaxis and chemokinesis are two common migratory behaviours in motile cells; the former involves directed movement along a chemical gradient, while the latter reflects speed changes based on chemical distribution~\cite{jin2017chemotaxis,ralt1994chemotaxis,son2016speed}. These behaviors may coexist, as in Myxococcus xanthus. Chemokinesis alone can lead to cell accumulation in low-motility zones, such as neutrophils near immune complexes. Similarly, self-propelled Pt/Au rods in opposing hydrogen peroxide and salt gradients accumulate in salt-rich, peroxide-poor regions due to chemokinesis~\cite{moran2021chemokinesis}. 
We believe that inverse chemokinesis, where motility controls chemical distribution, as highlighted in our study, can broaden the current understanding of migration dynamics in both natural and engineered contexts.

Finally, our findings have implications for synthetic active matter and artificial swarm intelligence, where minimal local interactions can generate coordinated behaviour on a larger scale~\cite{garnier2007biological}.  Microfluidic maze experiments~\cite{tweedy2020seeing}, or programmable colloidal systems~\cite{ Gompper2020}, can be used to test the predictions of our model and explore efficient navigation, aggregation, or dispersion under a minimal set of rules.

In summary, our work shows how biologically inspired local feedback mechanisms between motion and signaling can produce a wide array of collective structures. Though abstract by design, the model captures key ingredients relevant to both natural and artificial systems, offering a versatile platform to study emergent spatial organization in chemically interacting active agents.

\section*{Methods}
\label{sec:methods}
We employ a hybrid Monte Carlo simulation with $N$ agents guided by a self-generated chemical field. In this auto-chemotactic model, agents move in continuous time and space, while the chemical field - serving as spatial memory - is discretized on a triangular lattice (Fig. \ref{fig:scheme}(a)). Each agent updates its position as
$$
\mathbf{r}(t+1) = \mathbf{r}(t) + dr\, (\cos\theta(t), \sin\theta(t)),
$$
where $dr$ is sampled uniformly from $(0, v_0\Delta t)$, with $v_0\Delta t = 1$ fixed to represent the walker body size across all simulations.

The orientation $\theta(t)$ is determined by the local chemical environment, as illustrated in Fig. \ref{fig:scheme}(a$i$):
$$
\theta = \left< \theta_c \right> + \Delta \theta,
$$
where
$$
\la \theta_c \ra = \sum_{i=1}^{6} P(\theta_i|c_i)\, \theta_i,
$$
and $P(\theta|c) = \frac{c}{1 + c}$ is the probability of moving along direction $\theta$ given local concentration $c$. This biases motion toward higher concentrations, saturating at large $c$. The average direction $\left< \theta_c \right>$ (blue arrow in Fig. \ref{fig:scheme}(a$i$)) is perturbed by a random angle $\Delta \theta$ drawn from a Gaussian distribution with zero mean and standard deviation $\frac{\pi}{4}$ (depicted as an arc in the figure), forming a {\it chemical cone} with axis $\la \theta_c \ra$ and half-angle $\pi/4$. To account for excluded volume, the walker attempts a step $\Delta \mathbf{r}$ in direction $\mathbf{e}$ at speed $v_0$ with a certain probability:
$$P(\mathbf{r} \xrightarrow{} \mathbf{r+\Delta r}) =e^{-\frac{\Delta U}{k_BT}}$$ 
with  $\Delta U \equiv U(\mathbf{r}+ \Delta \mathbf{r},\mathbf{r'}) -U(\mathbf{r} ,\mathbf{r'})$, where, $U(\mathbf{r},\mathbf{r'})$ represents the Weeks-Chandler-Anderson (WCA) potential
$$
 U(r\equiv |\mathbf{r}-\mathbf{r'}|) =
\begin{cases}
    4\epsilon\left[\frac{1}{r^{12}} - \frac{1}{r^6} + \frac{1}{4}\right] ,& \text{if }r < 2^{1/6}\\
    0,              & \text{otherwise .}
\end{cases}
$$
 The chemical can evaporate stochastically with a  rate $\beta$. This leads to the chemical dynamics following:
$$ \frac{dc(\mathbf{r},t)}{dt} = \alpha(t) \sum_i \mathbf{\delta}(\mathbf{r},\mathbf{r}_i) - \beta c(\mathbf{r},t).$$
The deposition of the chemical is directly coupled to the particle dynamics. The rate $\alpha(t)$ of the chemical deposition at time $t$ depends on the agent's motility state and is defined as
\[
\alpha(t) =
\begin{cases}
    \alpha ,& \text{if }~d\mathbf{r}(t)\neq 0\\
    0,              & \text{otherwise.}
\end{cases}
\]
This rule links chemical deposition to agent motility: immobile particles, blocked by repulsion, do not deposit chemicals. Meanwhile, deposited chemicals evaporate over time (Fig. \ref{fig:scheme}(a$ii$)). Conversely, if a particle can move, it deposits an amount $\alpha\, dt$ on each surrounding chemical grid point (Fig. \ref{fig:scheme}(a$iii$)). Thus, in aggregates, mostly particles at the interface, where movement is possible, actively contribute to chemical deposition. \\

\vskip 0.5cm
\section*{Author contribution}
AC and DC designed the work. SSK performed the simulations and calculations. All the authors wrote the paper.

\section*{Data Availability}
All relevant data can be found within the article and the supplementary information.

\section*{Competing interests}  
There are no competing interests.

\section*{Acknowledgment}
DC acknowledges support from the Department of Atomic Energy (OM no. 1603/2/2020/IoP/R\&D-II/15028) and an ICTS-TIFR Associateship. AC acknowledges support from the Indo-German grant (IC-12025(22)/1/2023-ICD-DBT). Numerical computations were performed using the HPC facility at IISER Mohali, SAMKHYA (IOP Bhubaneswar), and PARAM Smriti. SSK acknowledges IOP Bhubaneswar for its hospitality during part of the research.
%

%\bibliographystyle{style}
%\bibliography{References}

%%%%%%%%%%%%%%%

%apsrev4-2.bst 2019-01-14 (MD) hand-edited version of apsrev4-1.bst
%Control: key (0)
%Control: author (8) initials jnrlst
%Control: editor formatted (1) identically to author
%Control: production of article title (0) allowed
%Control: page (0) single
%Control: year (1) truncated
%Control: production of eprint (0) enabled
%

%%%%%%%%%%%%%%

%%%%%%%%%% Merge with supplemental materials %%%%%%%%%%
\pagebreak
\widetext
\begin{center}
\textbf{\large Supplemental Materials: Self-organized fractal architectures driven by motility-dependent chemotactic feedback}
\end{center}
%%%%%%%%%% Merge with supplemental materials %%%%%%%%%%
%%%%%%%%%% Prefix a "S" to all equations, figures, tables and reset the counter %%%%%%%%%%
\setcounter{equation}{0}
\setcounter{figure}{0}
\setcounter{table}{0}
\setcounter{section}{0}
\setcounter{page}{1}
\makeatletter
\renewcommand{\theequation}{S\arabic{equation}}
\renewcommand{\thefigure}{S\arabic{figure}}
%\renewcommand{\bibnumfmt}[1]{[S#1]}
%\renewcommand{\citenumfont}[1]{S#1}
%%%%%%%%%% Prefix a "S" to all equations, figures, tables and reset the counter %%%%%%%%%%

\begin{widetext}
%\SItext
\section{Linear stability analysis}
In this section, we perform the linear stability analysis of the coupled equations for the evolution of particle density and chemical concentration (Eq. (2) in the main text). The equations are given as: 
\bea
\nonumber
    \dot{\rho} &=& -\gr \dt[\r \chi f'(c) \gr c-D \nabla \rho] \\
    \dot{c} &=& \alpha \rho (\r_c-\r)-\beta c  
    \label{eq:rho,c}
\eea
This analysis is performed about the homogeneous state $(\rho_0, c_0)$. The homogeneous chemical concentration is obtained by putting $\dot{c} = 0$, giving $c_0 = \rho_0(\rho_c - \rho_0)\alpha/\beta$. Substituting $\r =\r_0+\delta \r, c=c_0+\delta c$ in Eq.~\ref{eq:rho,c} where $(\delta\r, \delta c)$ denote perturbations around the homogeneous state, we get 
\begin{equation}
    \begin{split}
      \frac{\partial \delta \rho}{\partial t} & =-\nabla \cdot\left[\chi \rho_0 f^{\prime}(c) \nabla \delta c- D \nabla \delta \rho\right] \\
     & = -\chi \rho_0 f^{\prime}(c) \nabla^2 \delta c + D \nabla^2 \delta \rho \\   
     \frac{\d \delta c}{\d t} &= \a (\r_c - 2 \r_0 ) \dr - \beta \dc + D_c \gr^2\dc \\
    \end{split}
    \label{rhol}
\end{equation}
Expanding in Fourier modes, $\dr = \dr_0 e^{i\mathbf{q}.\rb +\omega t}, \dc = \dc_0 e^{i\mathbf{q}.\rb +\omega t} $
we get, 
\begin{equation}
    \omega \begin{pmatrix}
        \dr_0 \\ \dc_0
    \end{pmatrix} =\begin{pmatrix}
\A & \B\\
\C& \D
\end{pmatrix} 
\begin{pmatrix}
        \dr_0 \\ \dc_0
    \end{pmatrix}
\end{equation}
where 
\begin{equation*}
\begin{split}
    \A& = -D q^2, ~~ \B = q^2 \chi \r_0 f'(c_0), \\ 
    \C& =\a(\r_c - 2 \r_0 ), ~~
    \D = -\b .\\
\end{split}    
\end{equation*} 
The characteristic equation is 
$\omega^2 - \omega\,\textrm{Tr} + \textrm{Det} = 0$ with $\textrm{Tr} = ( \A+\D)$ and $\textrm{Det} = \A\D- \B\C$. It has solutions
%$\omega^2 - \l( \A+\D\rt)\omega + \A\D- \B\C =0$
%gives: 
\begin{align}
\omega_{\pm} = \frac{1}{2} [\textrm{Tr} \pm \sqrt{\Delta}].
%,~{\rm with~ discreminant}~ \Delta=Tr^2-4\, Det. %\\
% \omega_{\pm} = \frac{1}{2} \l [ \A + \D \pm \sqrt{\l (\A - \D\rt )^2 + 4 \B\C} \rt ].
\end{align}
where $\Delta=\textrm{Tr}^2-4\, \textrm{Det}.$ Note that $\textrm{Tr} < 0$. Therefore,  instability arises when $\textrm{Det} < 0$:
\begin{gather}
    \textrm{Det} = \A\D - \B\C < 0 \notag \\    
      q^2(\b D - \a(\r_c - 2 \r_0 )\chi \r_0 f'(c_0)) < 0 \notag \\
      \b D - \a(\r_c - 2 \r_0 )\chi \r_0 f'(c_0) < 0 \notag \\
     \frac{\b}{\a} D - \frac{\a(\r_c - 2 \r_0 )\chi \r_0}{(1+c_0)^2}  < 0
\end{gather}
As noted earlier, $\dot{c} = 0$, gives ${\a}/{\b} = {c_0}/{\r_0(\r_c - \r_0)}$. The instability condition then becomes 
\bea
c_0^2 + (2-\chi')c_0 + 1 < 0
\label{eq::Instability_quadratic}
\eea 
where $\chi' = \frac{(\r_c - 2 \r_0 )\chi }{(\r_c - \r_0 )D }$. Solving Eq. \ref{eq::Instability_quadratic} gives
\begin{gather}
    C_{-} < c_0 <C_{+} \notag \\    
    i.e. \,\, C_{-} < \frac{\a \r_0(\r_c - \r_0)}{\b}  < C_{+}
\end{gather}
where $C_{\pm}= \frac{\chi' -2 \pm \sqrt{\chi'^2 - 4\chi'} }{2}$. It is assumed that $\chi' > 4$ for the existence of real solutions. Note that $C_{\pm}$ denotes the range of values of $\alpha/\beta$ where the homogeneous state becomes unstable, leading to pattern formation. 

\section{Numerical Integration}

\begin{figure}[t]
    \centering
    \includegraphics[width=\textwidth]{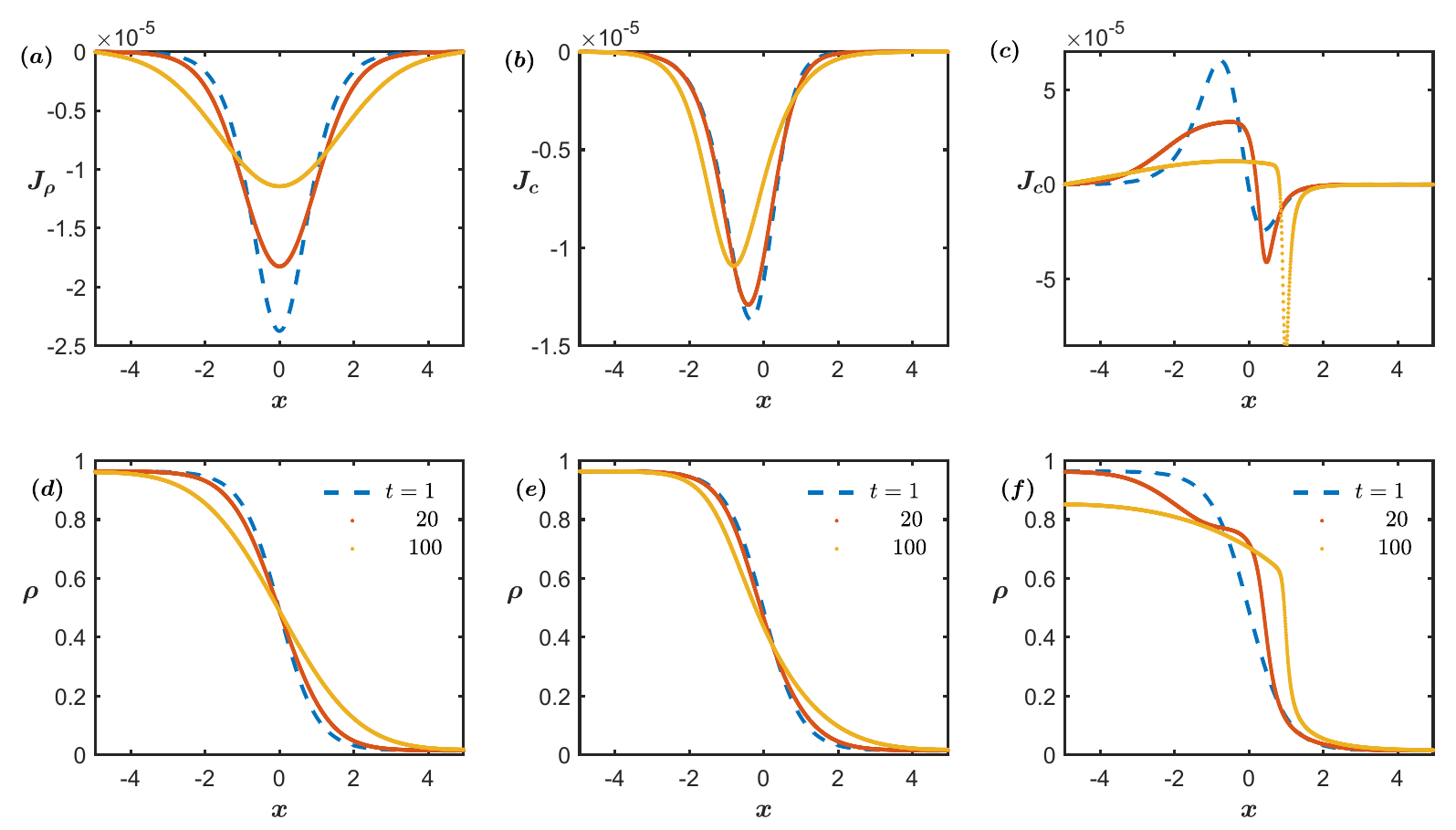}
    \caption{
    Time evolution of particle density $\rho$ and associated currents from numerical integration of Eq.~\ref{eq:rho,c} for: (a) $J_\rho = -D\nabla \rho$ without chemical, (b) $J_c = \chi \rho \nabla f(c)$ at a constant chemical deposition, and (c) $J_c$ using density-dependent deposition $\alpha \rho(1 - \rho/\rho_c)$. The initial density profile is $\rho(x) = \frac{1}{2}(1 - \tanh(x)\,)$, with chemical fields set to their respective steady states. In (b), chemotactic drift competes with diffusion, resulting in slower spreading. In (c), density-dependent deposition drives front propagation. 
    Density evolution for (a), (b), and (c) is shown in (d), (e), and (f), respectively. Profiles are shown at times $0$, $20\tau$, and $100\tau$, as indicated in the figures.
    Parameters: $D = 0.01$, $\chi = 1.0$, and $\alpha/\beta = 0, 2, 20$ for (a–c). 
       }
    \label{fig:S1}
\end{figure}
In our model, in the main text, we used a mobility dependent chemical deposition. To compare it with simpler models, where deposition does not depend on mobility, here we perform numerical integrations of the relevant equations for the evolution of the particle density, $\r$ and the chemical $c$. For this purpose, we use two different chemical evolution equations - the first is as in Eq.~\ref{eq:rho,c}, while the second set of equations is given as:
\bea
\nonumber
    \dot{\rho} &=& -\gr \dt[\r \chi f'(c) \gr c-D \nabla \rho] \\
    \dot{c} &=& \alpha \rho -\beta c. 
    \label{eq:rho,c1}
\eea
For this purpose, we start with the same initial density profile
$\r(x) = (\rho_i/2)(1 - \tanh{(x/w)})$, with $\rho_i = 0.95$ and $w = 1$. To integrate the equations in one dimension, we used a finite difference scheme with $\sigma, \tau$ as units of length and time, respectively. The discretization length, $dx =0.005\sigma$ and the time interval between each step are chosen to be $dt =0.0002\tau$, keeping Courant-Friedrichs-Lewy (CFL) condition for stability into consideration.

In Fig.~\ref{fig:S1}(a) we look at the diffusive current $J_{\r} = -D\nabla \r$ for $\frac{\a}{\b} = 0$ at three different time points. This serves as the control. As expected, the dynamics is purely diffusive and shows a smooth density profile with particles moving from $x < 0$ to $x > 0$ as time progresses (Fig.~\ref{fig:S1}(d)). Next, we solve Eq.~\ref{eq:rho,c1} for $\frac{\a}{\b}  = 2$. In this case the chemical deposition is independent of the mobility of the particles. There is a competition between the diffusive flux $J_{\r}$ and the chemical drive $J_c = \chi \rho \nabla f(c)$. The spatial variation of $J_c$ is plotted in Fig.~\ref{fig:S1}(b) at three different time points. The resultant density profile now shows a much slower time evolution (see Fig.~\ref{fig:S1}(e)) as compared to the purely diffusive case ($\alpha = 0$, Fig.~\ref{fig:S1}(d)). 

Finally, we look at the case where the deposition is mobility dependent and solve Eq.~\ref{eq:rho,c} for $\frac{\a}{\b} = 20$. The spatial variation of  $J_c$ is now very different from that of Fig.~~\ref{fig:S1}(b). Due to the density dependence, deposition starts decreasing with increase in density near $\rho = 0.5$ leading to a maximum in $J_c$.  For $x < 0$, there is a positive chemical drive as shown in Fig.~\ref{fig:S1}(c). This drive pushes particles  from $\r > 0.5$ towards the lower dense region and results in a local peak near $\r \approx  0.5$ as shown in ~Fig.\ref{fig:S1}(f). Further, this peak is stabilized by the chemical and also the front propagates towards $x > 0$ as time progresses shown in ~Fig.\ref{fig:S1}(f). 

We can therefore conclude that mobility dependent chemical deposition serves as a key factor for the instability towards network formation. In Fig.~\ref{fig:S2}, we show the simulation data for chemical concentration versus particle density at steady state at two distinct regions of the phase diagram: (a) DN phase and (b) SC phase. In both the phases, the corresponding fit to a functional dependence of the chemical concentration of the form $c = a\r - b\r^2$ shows a good agreement. Note that this functional form is as expected from Eq.~\ref{eq:rho,c} with $\dot{c} = 0$.

\begin{figure}[tbh!]
    \centering
    \includegraphics[width=\textwidth]{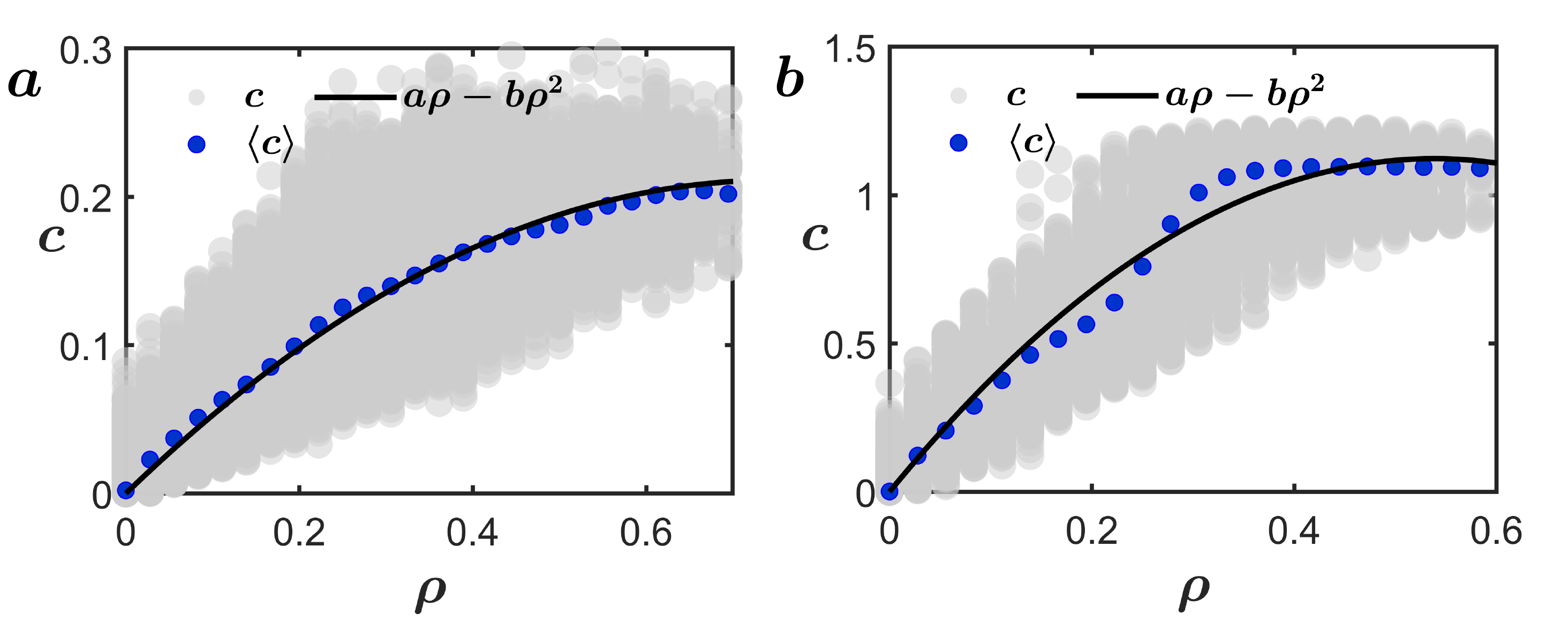}
    \caption{Plot of local particle density versus chemical concentration (scaled by $\frac{\alpha}{\beta}$) at steady state. Panel (a) shows the fractal network phase (DN, $\alpha = 5$, $\beta = 0.008$), while (b) shows the aggregate SC phase ($\alpha = 100$, $\beta = 0.008$). The solid line fits the non-monotonic function $f(\rho) = a\rho - b\rho^2$, with $(a, b) = (0.56 \pm 0.02, 0.38 \pm 0.03)$ in (a) and $(4.2 \pm 0.3, 3.9 \pm 0.5)$ in (b). Thus $\r_c = a/b$ is $1.47$ in (a) and 1.07 in (b).}
    \label{fig:S2}
\end{figure}

\newpage
\section{Cluster size distribution}
\begin{figure}[tbh!]
    \centering
    \includegraphics[width=1.1\textwidth]{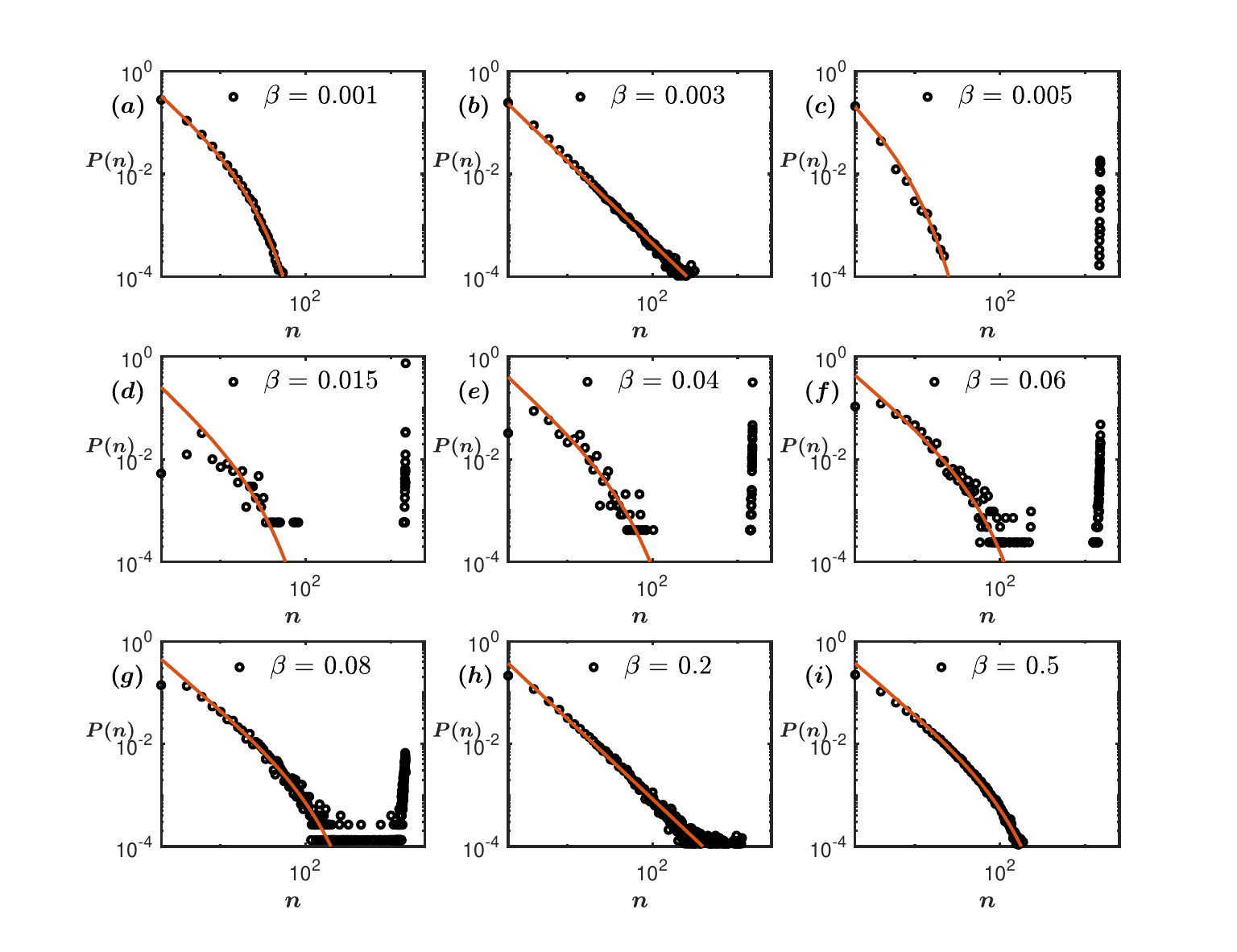}
    \caption{~(a-i) Cluster size distribution, $P(n)$ at $\a=75.0$. The data is fitted with the functional form: $P(n) \sim \exp(-n/n^*)/n^\nu $.
    %[$ p\exp(-qx)/x^{r}$]. 
    We have fixed $\nu \approx 1.5$ as in \cite{Majumdar1998,Peruani2013}. From (a) to (i), the fitting parameter $1/n^*$ is given by 0.067,0, 0.167, 0.05, 0.028, 0.025, 0.01, 0, 0.011. The vanishing $1/n^*$ in (b) and (h) marks the phase boundaries of the homogeneous phase.
     } 
     \label{fig:SProb_dist}
\end{figure}
In Fig.~\ref{fig:SProb_dist}, we show the variation of the cluster size distribution, $P(n)$ with $\beta$ at an intermediate $\alpha$ value.  In the top row, we show the variation of the distribution as we move from the RHP phase to the SC phase.  % $\beta = 0.001$ corresponds to the RHP phase (Fig.~\ref{fig:SProb_dist}(a)).
In Fig.~\ref{fig:SProb_dist}(a) we see a distribution of the form $P(n) \sim \exp(-n/n^*)/n^\nu $ corresponding to the RHP phase. In the SC phase, the distribution in addition to this functional form shows a large aggregate near $n \approx N$ as expected. At the boundary between the RHP and SC phase, the distribution shows a power-law dependence. 

In the middle row, we move across values of $\beta$ corresponding to the boundary between SC and DN phase (Fig.~\ref{fig:SProb_dist}(d)), deep in the DN phase (Fig.~\ref{fig:SProb_dist}(e)) and at the boundary between the DN and SN phase (Fig.~\ref{fig:SProb_dist}(f)). All these plots show a large aggregate near $n \approx N$ with increasing spread of the distribution.  

In the bottom row, we show the variation of the distribution as we go from the SN phase to the HP phase. Inside the SN phase (Fig.~\ref{fig:SProb_dist}(g)), there is a spread in the distribution, indicating the formation of a small network of clusters. Again, near the boundary between the SN and HP phase we see a power-law distribution (Fig.~\ref{fig:SProb_dist}(h)). With increasing $\beta$, we enter the homogeneous HP phase with an exponential distribution of cluster sizes (Fig.~\ref{fig:SProb_dist}(i)).

\begin{figure}[tbh!]
    \centering
    \includegraphics[width=0.5\textwidth]{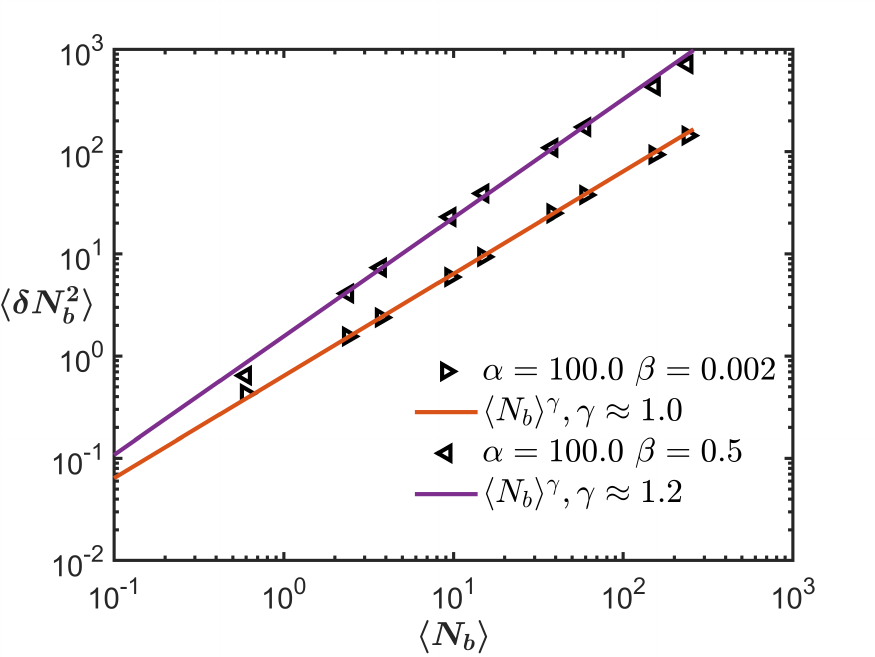}
    \caption{~ Plot of giant number fluctuations, $\la \delta N_b^2 \ra =  \langle N_b^2 \rangle -  \langle N_b \rangle^2$ versus $\langle N_b \rangle$, where $N_b$ denotes the number of particles and  $\langle \cdots\rangle$ denotes averages over different boxes of linear dimension $\approx 2,4,5,8,10,16,20,32,40$ in units of $r_c$. $\la \delta N_b^2\ra \sim \la N_b \ra^\gamma$ with $\gamma \sim 1.2$ suggesting giant number fluctuation in HP. } 
    \label{fig:gnf}
\end{figure}
The distributions clearly indicate homogeneous phases at two regimes, one at high $\b$ (Fig.~\ref{fig:SProb_dist}(a)) corresponding to HP phase and another at very low $\b$ (Fig.~\ref{fig:SProb_dist}(b)), corresponding to RHP phase. However, there are clear differences between the HP and RHP phases. This is due to the presence of giant number fluctuations in HP (Fig.~\ref{fig:gnf}), which is absent in RHP. As a result, the exponential cluster size distribution in HP requires a necessary algebraic correction in the form $n^{-\nu}$ multiplying $\exp(-n/n^*)$. This algebraic correction is not essential to describe RHP (Fig.~\ref{fig:SProb_dist}).  
\begin{figure}[tbh]
    \centering
    \includegraphics[width=0.5\textwidth]{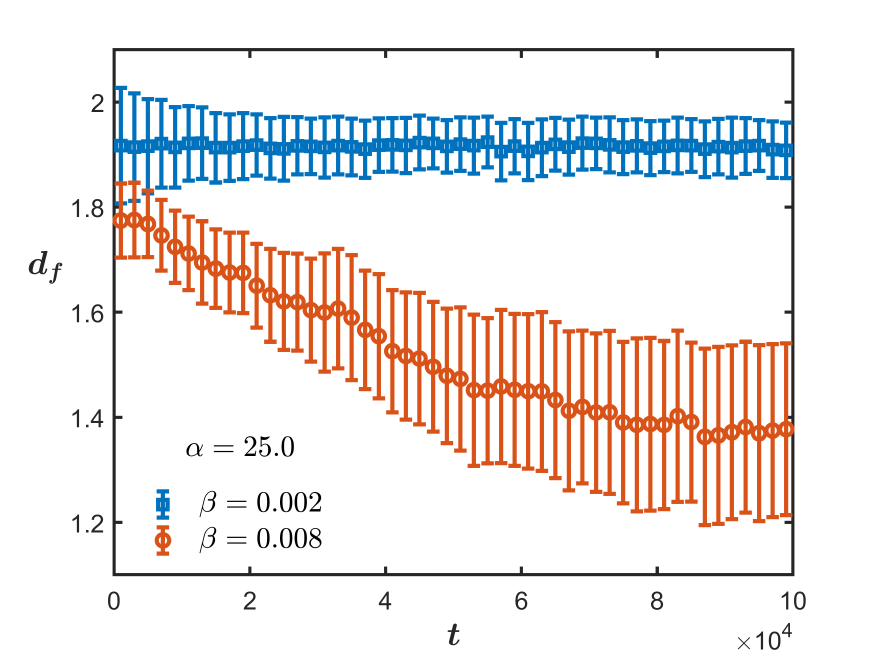}
    \caption{Evolution of the fractal dimension $d_f$ following a quench from a compact configuration to steady state at $\alpha = 25.0$ for two decay rates: $\beta = 0.002$ (SC phase) and $\beta = 0.008$ (in DN phase). For the evolution of pattern to network of filamentous elements, $d_f$ decreases toward 1,  while during the evolution to homogeneous structure, $d_f$ remains close to 2.}
    \label{fig:fractal}
\end{figure}

\section{Fractal characterization}

We quantify the fractal nature of the network phase and also distinguish the non-network structure using the concept of correlation dimension \cite{Grassberger1983,strogatz}, which calculates the total number of pairs of particles which have a distance between them that is less than a certain distance. For this purpose, a particle, $i$, is chosen, and $N_i(r)$, the number of particles within a distance, $r$ around $i$, is calculated. 
%This gives us the pointwise dimension as we vary $r$. To get the overall dimension,
$N_i(r)$ is averaged over all the particles to obtain the correlation dimension, $C(r)$. For a fractal, this is expected to show a power law behavior with the exponent giving the fractal dimension $d_f$:
\begin{equation}
    C(r) \sim r^{d_f}.
    \label{frac}
\end{equation}

In Fig.~\ref{fig:fractal}, we show the time evolution of the fractal dimension obtained from Eq.~\ref{frac} for two values of $\beta$ corresponding to the DN and SC phases, starting as before from a quenched system. In the quenched state, the particles are packed in a square box with $d_f = 2$, the system dimension. In the SC phase, $d_f$ fluctuates about the same, indicating a homogeneous aggregate. In the DN phase, $d_f$ decreases sharply with time before saturating at a lower value $~\sim 1.4$, characterising a network structure.

\section{System span}
The system span, $s$ can be used to estimate the area covered by a system of connected particles.
In Fig. \ref{fig:convex_hull} we show the system span, $s$, for the SC phase in ~Fig. \ref{fig:convex_hull}(a) and the DN phase in ~Fig. \ref{fig:convex_hull}(b), which we calculate using the convex hull method. We start with the quenched system, i.e., all particles clustered at the centre of the simulation box. After $10^6$ steps of evolution, we measure the span of this cluster. We use the convhull function in MATLAB, which uses the Qhull algorithm\cite{MATLAB}. This gives a polygon marked in red as in the ~Fig. \ref{fig:convex_hull}, and $s$ is obtained by calculating the area of the polygon, $A$ scaled by the simulation box area:  $s = {A}/{L_xL_y}$. 
\begin{figure}[tbh!]
    \centering
    \includegraphics[width=1.0\textwidth]{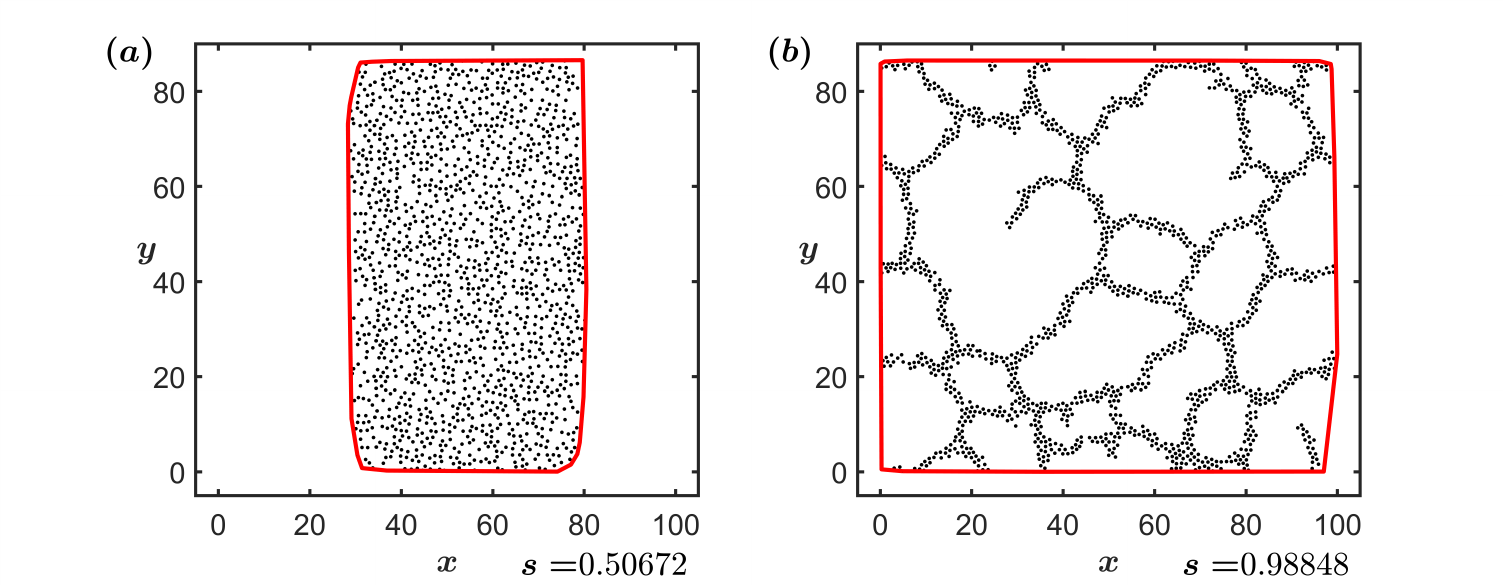}
    \caption{~ System span, $s$ of the (a) SC phase and (b) DN phase using the convex hull method. $s$ is lower in the SC phase and has a high value in the DN phase, quantifying the system spanning network structure.} 
    \label{fig:convex_hull}
\end{figure}

\section{Growth laws}
In Fig.\ref{fig:clustersizevst}(a) we show mean cluster size growth laws. A quench to the SC phase shows growth law $\la \bar n \ra \sim t^{1/z}$ consistent with $z \approx 3$ for the conserved dynamics (for $t> 10^4$). In contrast, a quench to the DN phase, despite showing initial consistency with such conserved dynamics, quickly crossovers to a much faster exponential growth $\la n \ra \approx n_{\rm sat} - n_0 \exp(-t/t^\ast)$ where $n_{\rm sat}\approx 1474$, $\ n_0 \approx 1665$ and $t^\ast \approx 46530$. Arguably, $t^\ast$ will grow with system size.    
\begin{figure}[tbh!]
    \centering
    \includegraphics[width=0.5\textwidth]{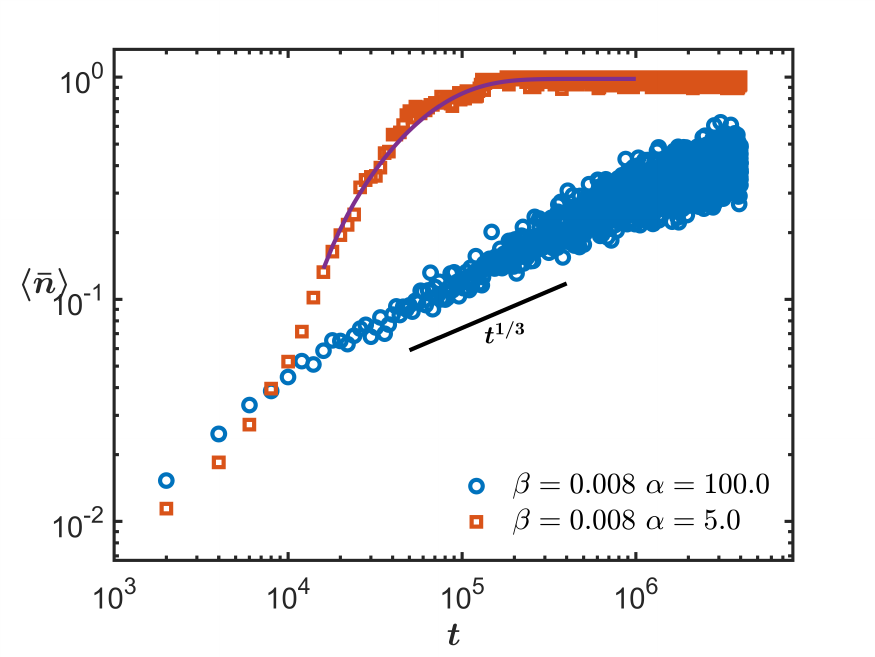}
    \caption{~ Evolution of mean cluster size,  $\langle n(t) \rangle$ scaled with the total number of particles(N), $\la \bar{n} \ra=  \frac{\langle n(t) \rangle}{N}$ averaged over 20 initial configurations.The purple line represents $\bar n_{\rm sat} - \bar n_0 \exp(-t/t^\ast)$ with $\bar n_{\rm sat} \approx 0.98, \bar n_0 \approx 1.2$ and $t^\ast \approx 46530$ } 
    \label{fig:clustersizevst}
\end{figure}

\section{Perturbation study}
\begin{figure}[tbh!]
    \centering
\includegraphics[width=1.0\textwidth]{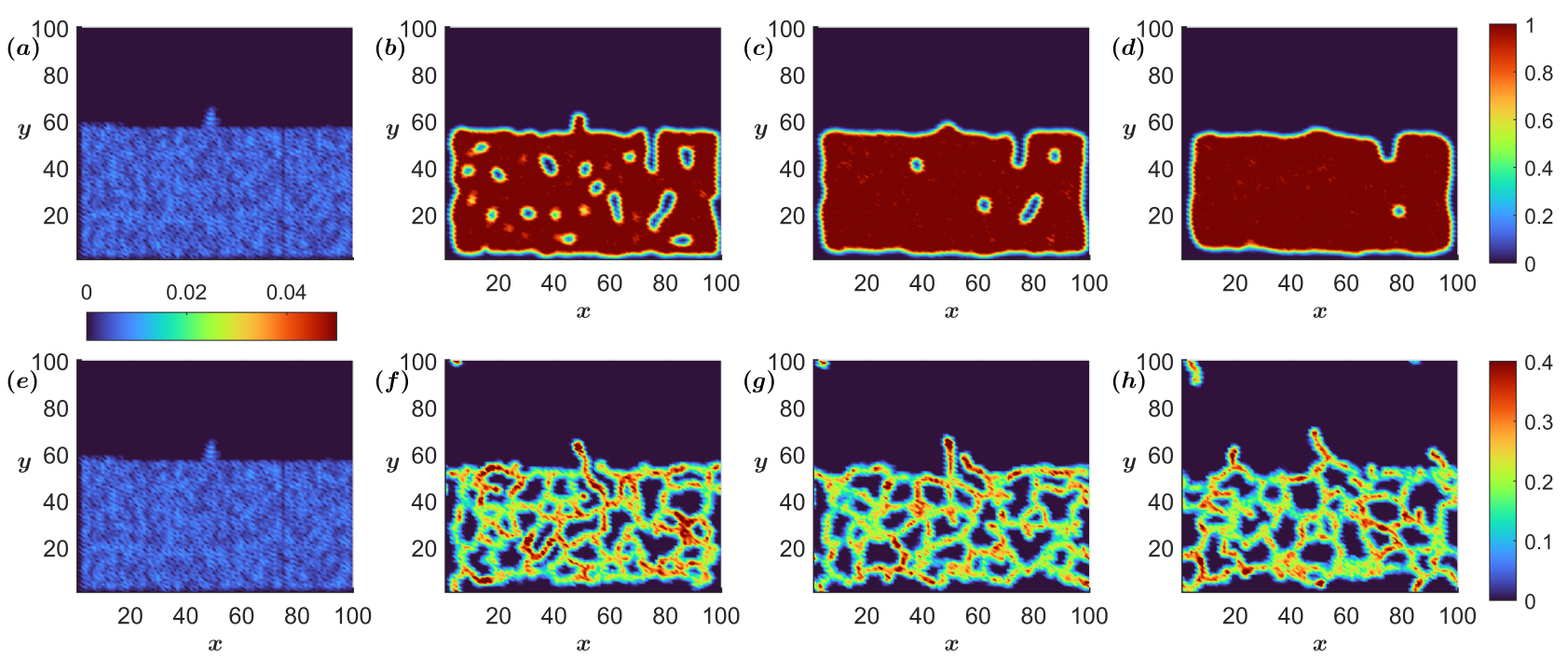}
    \caption{~Evolution of the perturbed state. (a,e) present the initial chemical configuration with the concentration shown in the colorbar between them. (b-d) show the evolution of the chemical configurations of the SC phase at $\a=25, \b=0.002$ and DN phase in (f-h) at $\a=25, \b=0.008$. Configurations are at timesteps=$1,4000,8000,16000$ for columns $1-4$ respectively. The colorbar indicates the chemical concentration scaled with $\frac{\a}{\b}.$ %for (b-d) at the top and (f-h) at the bottom.
    A separate colorbar for chemical concentration is used for (a) and (e).
    It is seen that the perturbation is suppressed in the SC phase whereas it grows in the DN phase.}  
    \label{fig:Perturb}
\end{figure}

We started with a state where the particles were arranged uniformly with $0<y<50$, but with a small sinusoidal perturbation at the centre, $x\approx 50$, as shown in Fig.~\ref{fig:Perturb}(a,e). The system is then allowed to evolve at parameter values corresponding to the SC phase ($\a=25, \b=0.002$) and the DN phase ($\a=25, \b=0.008$). The time evolution of the chemical configurations are presented in the top (b-d) and bottom (f-h) panels in Fig.~\ref{fig:Perturb}. The evolution is different for both cases. In the SC phase, the perturbation is suppressed in time with the formation of a macrocluster. In the DN phase, the perturbation grows as time progresses. This is consistent with the analytical picture presented before. 

\section{Pair correlation}
\begin{figure}[tbh!]
    \centering
    \includegraphics[width=0.5\textwidth]{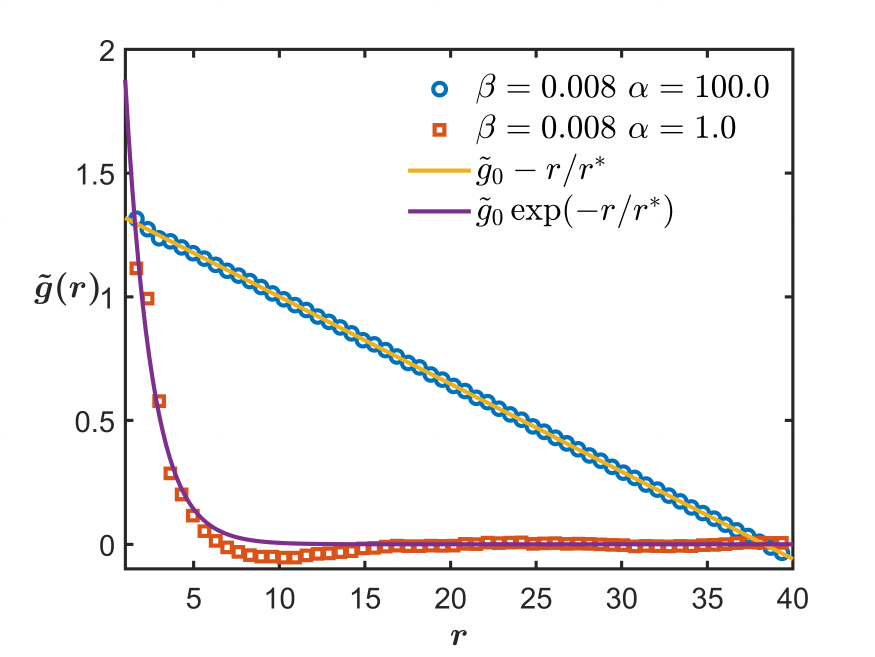}
    \caption{~Pair correlation function, $\tilde g(r) = g(r)-1$ at $\b=0.008$ for two different phases SC phase at $\a =100.0$ and DN phase at $\a =1.0$. The functions used to fit are mentioned in the plot. For the linear function $\tilde g_0=1.35, r^*= 28.29$; and exponential function has $ \tilde g_0 =3.56, r^*=1.55$ } 
    \label{fig:g_r}
\end{figure}
In the DN phase, the pair correlation decays in an exponential manner. This behavior is similar to what is seen in an equilibrium fluid. It suggests that the structure is relatively uniform and lacks sharp features.

In contrast, the SC phase shows a different trend. Here, the pair correlation decays linearly. This matches the prediction of Porod's law. It indicates the presence of sharp interfaces between regions.

\section{Supplementary movies}

Movie-1: {RHP and HP phase at $\a =100.0, \b =0.002$ and $\a =100.0, \b =0.5$ respectively.}

Movie-2: {SN phase at $\a =1.0, \b =0.1$ }

Movie-3: {DN phase at $\a=0.01,\b=0.001$}

Movie-4: {SC phase at $\a =100.0, \b =0.008$ }

Movie-5: {Quench Dynamics corresponding to Fig. 3 in the main text}

\end{widetext}

\end{document}